\documentstyle[11pt,newpasp,twoside,epsf]{article}

\begin{document}

\title{Dark Matter in Dwarf Galaxies: Latest Density Profile Results}
\author{Joshua D. Simon, Alberto D. Bolatto, Adam Leroy, and Leo Blitz}
\affil{Department of Astronomy, University of California at Berkeley, 
601 Campbell Hall, CA  94720}

\begin{abstract}
We present high-resolution two-dimensional velocity fields in
H$\alpha$ and CO of the nearby dwarf galaxy NGC 2976.  Our
observations were made at both higher spatial resolution ($\sim75$ pc)
and higher velocity resolution (13 km~s$^{-1}$ in H$\alpha$ and 2
km~s$^{-1}$ in CO) than most previous studies.  We show that NGC 2976
has a very shallow dark matter density profile, with $\rho(r)$ lying
between $\rho \propto r^{-0.3}$ and $\rho \propto r^{0}$.  We
carefully test the effects of systematic uncertainties on our results,
and demonstrate that well-resolved, two-dimensional velocity data can
eliminate many of the systematic problems that beset longslit
observations.  We also present a preliminary analysis of the velocity
field of NGC 5963, which appears to have a nearly NFW density profile.
\end{abstract}

\section{Introduction}

It is well-known by now that there is a substantial disagreement
between the observed dark matter density profiles of many dwarf and
low-surface brightness galaxies and the density profiles predicted by
numerical Cold Dark Matter (CDM) simulations (e.g., Flores \& Primack
1994; Burkert 1995; Navarro, Frenk, \& White 1996, hereafter NFW;
Moore et al. 1999).  The significance of this disagreement, though,
remains controversial.  A number of authors attribute the problem to
failures of the simulations, or of the CDM model itself (de Blok et
al. 2001a; de Blok, McGaugh, \& Rubin 2001b; Borriello \& Salucci
2001; de Blok, Bosma, \& McGaugh 2003), while others argue that
systematic uncertainties in the observations make such conclusions
premature (van den Bosch et al. 2000; van den Bosch \& Swaters 2001;
Swaters et al. 2003, hereafter SMVB).

We address this controversy with a new study that combines a number of
techniques to overcome the systematics in the observations.  We are
acquiring very high-quality data on a limited sample of galaxies to
investigate the importance of systematic effects in detail.  The
results of this study should make clear whether systematic problems in
the data are at fault, or whether there actually is a fundamental
conflict between the theory and the observations.

Our program includes (1) two-dimensional velocity fields obtained at
optical (H$\alpha$), millimeter (CO), and centimeter (H {\sc i})
wavelengths, (2) high angular resolution ($\sim5\arcsec$), (3) high
spectral resolution ($\la 10$ km~s$^{-1}$), (4) multicolor optical and
near-infrared photometry, and (5) nearby dwarf galaxies as targets.
The combination of these features greatly reduces our vulnerability to
systematic uncertainties (see Simon et al. 2003).

In two previous papers, we reported on rotation curve studies of the
dwarf spiral galaxies NGC~4605 and NGC~2976 (Bolatto et al. 2002;
Simon et al. 2003).  In this paper we highlight some of those results,
focusing on tests for systematic uncertainties in the NGC~2976 data
set, and we also present preliminary results on the third galaxy in
our study, NGC~5963.

\section{The Dark Matter Halo of NGC~2976}
\label{ngc2976}

NGC~2976 is a regular Sc dwarf galaxy located in the M81 group at a
distance of 3.45 Mpc.  The galaxy has absolute magnitudes of M$_{B} =
-17.0$ and M$_{K} = -20.2$, and a total mass of $3.5 \times 10^{9}
M_{\odot}$; it is somewhat less luminous and less massive than the
Large Magellanic Cloud.  In optical and near-infrared images it is
clear that NGC~2976 is a bulgeless, unbarred, pure disk system, which
makes it an ideal galaxy for mass modeling.

\subsection{Observations}
\label{2976observations}

\begin{figure}[!t]
\plotone{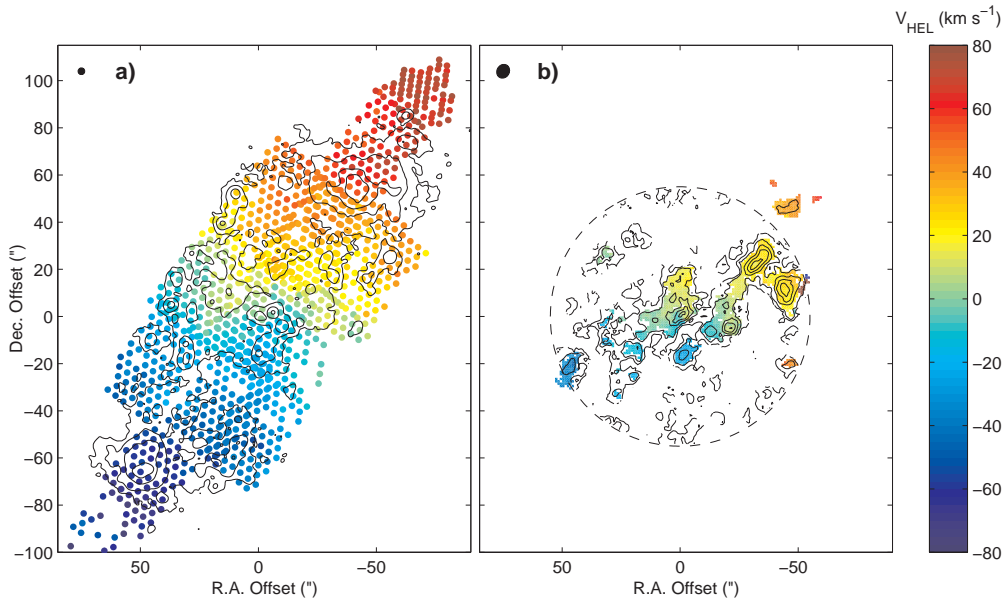}
\caption{(a) H$\alpha$ velocity field of NGC~2976 from WIYN
observations.  The contours represent integrated H$\alpha$ intensity.
(b) CO velocity field from BIMA observations.  The contours represent
integrated CO intensity.  The angular resolution of each dataset is
shown in the upper left corners.  
\label{velfield}}
\end{figure}

Our observations of NGC~2976 include a two-dimensional H$\alpha$
velocity field (obtained with a multi-fiber spectrograph on the WIYN
telescope), a two-dimensional CO velocity field (obtained with the
BIMA interferometer), multicolor optical photometry, and near-infrared
2MASS imaging.  The velocity fields are both shown in Figure
\ref{velfield}.  For details of the data reduction and analysis, see
Simon et al. (2003).

\subsection{Rotation Curve}
\label{rcsec}

\begin{figure}[!t]
\plotone{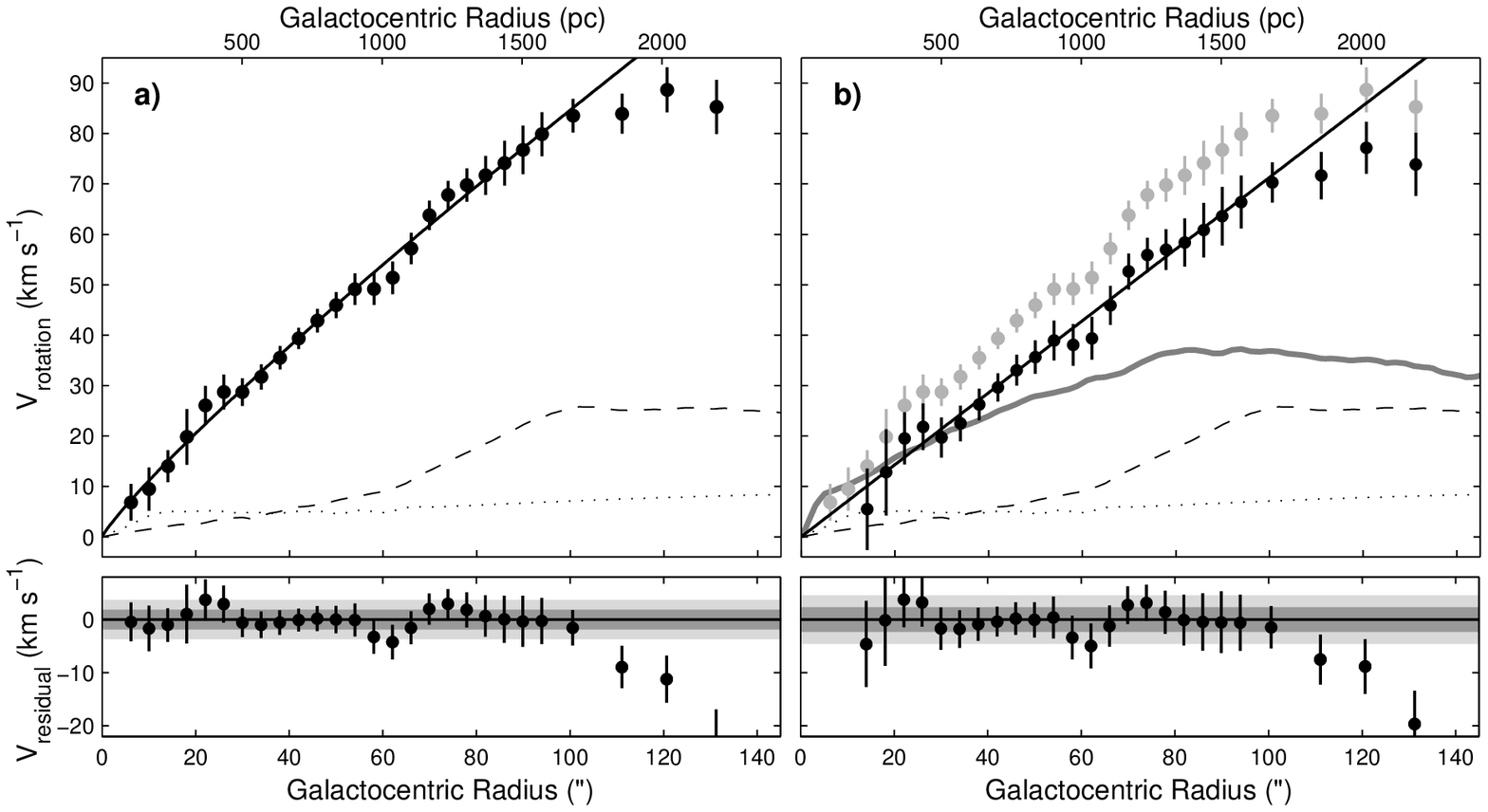}
\caption{(a) Minimum disk rotation curve of NGC~2976.  The observed
rotation velocities are plotted as black circles and the error bars
are combined statistical and systematic uncertainties.  The dashed and
dotted curves represent the rotation velocities due to H {\sc i} and
H$_{2}$, respectively.  A power law fit to the rotation curve is shown
by the solid black curve.  The corresponding density profile is $\rho
\propto r^{-0.27}$.  Residuals from the fit are displayed in the lower
panel, and $1\sigma$ and $2\sigma$ departures from the fit are
represented by the shaded regions.  (b) Maximum disk rotation curve of
NGC~2976.  The stellar disk (thick gray curve) is scaled up as high as
the observed rotation velocities (gray circles) allow.  The stellar
disk shown here has $M_{*}/L_{K} = 0.19$ \mbox{$M_{\odot}/L_{\odot K}$}.
After subtracting the rotation velocities due to the stars and the
atomic and molecular gas from the observed rotation curve, the dark
matter rotation velocities are displayed as black circles.  The two
missing data points near the center of the galaxy had $v_{rot} <
v_{*,rot}$, yielding imaginary $v_{halo}$.  The solid black line is a
power law fit to the halo velocities (for $14\arcsec < r <
105\arcsec$) which corresponds to a density profile of
$\rho_{\mbox{{\tiny DM}}} \propto r^{-0.01}$.  The halo residuals
after the power law fit are displayed in the bottom panel.
\label{ngc2976rc}}
\end{figure}

We used various tilted-ring modeling algorithms to convert the
observed velocity field into a rotation curve (see Simon et al. 2003).
The rotation curve of NGC~2976 is well-described by a single power law
from the center of the galaxy out to a radius of almost 2 kpc, as
displayed in Figure \ref{ngc2976rc}a.  The rotation curve only begins
to deviate systematically from power law behavior at $r \approx
110\arcsec$ (1.84 kpc).  The \emph{total} (baryonic plus dark matter)
density profile corresponding to the rotation curve is
$\rho_{\mbox{{\tiny TOT}}} = 1.6 (r/1 \mbox{ pc})^{-0.27 \pm 0.09}
M_{\odot}$~pc$^{-3}$.  Because the baryons are almost certainly
centrally concentrated, this density profile represents the cuspiest
possible shape for the dark matter halo.

\subsection{Dark Matter Density Profile}
\label{dmlimits}

To explore the full range of possible dark matter density profile
slopes, we must account for the rotational velocities contributed by
the baryonic components of the galaxy.  Because our images of NGC~2976
do not reveal a bulge or a bar, and its nucleus is dynamically
unimportant, the only relevant reservoirs of baryons to consider are
the stellar and gaseous disks.  We calculated the rotation curves due
to the stars and gas directly from the observed surface density
profiles, assuming an infinitely thin disk.  The only free parameter
in these calculations is the K-band mass-to-light ratio of the stellar
population ($M_{*}/L_{K}$).

Under the assumption that the density profile can be described with a
power law, $\rho_{\mbox{{\tiny DM}}} \propto r^{-\alpha_{\mbox{{\tiny
DM}}}}$, we found that if $\mbox{$M_{*}/L_{K}$} > 0.19
\mbox{$M_{\odot}/L_{\odot K}$}$, $\alpha_{\mbox{{\tiny DM}}} < 0$ and
the density of the dark matter halo is increasing with radius.  Such a
dark matter configuration is probably unphysical, so we consider 0.19
\mbox{$M_{\odot}/L_{\odot K}$} to be a firm upper limit to the stellar
disk mass-to-light ratio, with the corresponding lower limit to
$\alpha_{\mbox{{\tiny DM}}}$ of 0.  Because of the extremely low value
of the maximal disk mass-to-light ratio, the galaxy must contain an
essentially maximal disk.  The dark matter density profile for the
maximal disk is $\rho_{\mbox{{\tiny DM}}} = 0.1 (r/1\mbox{ pc})^{-0.01
\pm 0.13} M_{\odot}\mbox{ pc}^{-3}$.  This represents the most likely
shape for the dark matter halo.  Since the slope of the total density
profile of the galaxy represents the absolute upper limit for the
slope of the dark matter density profile, we conclude that the dark
matter density profile is bracketed by $\rho_{\mbox{{\tiny DM}}}
\propto r^{-0.27 \pm 0.09}$ and $\rho_{\mbox{{\tiny DM}}} \propto
r^{0}$.

\section{Systematics}
\label{systematics}

Several authors have recently discussed the systematic uncertainties
that can significantly alter observed rotation curves.  SMVB used
simulated observations to argue that systematic effects alone could
account for the difference between the predicted and observed density
profile slopes.  de Blok et al. (2003) carried out similar
simulations, but concluded that the systematic effects were not strong
enough to cause cuspy density profiles to appear to have cores.  We
have shown that, \emph{for the specific case of NGC~2976}, the known
systematic problems do not distort the rotation curve (Simon et
al. 2003; this work).  The worst systematics can probably be minimized
or avoided \emph{in general} by using two-dimensional velocity fields
and by making velocity measurements at very high precision
(Blais-Ouellette et al. 1999; van den Bosch \& Swaters 2001; Bolatto
et al. 2002; SMVB), but additional observations confirming this
expectation for more galaxies are needed.

\subsection{Rotation Curve Fitting Systematics}
\label{rcsystematics}

Because of the high precision of our velocity measurements, the
statistical uncertainties on both the rotation curve and the radial
velocity curve are negligible (less than 1 km~s$^{-1}$ everywhere).
Therefore, the limiting factors on the accuracy of the rotation curve
are the systematic uncertainties associated with our fit.

\subsubsection{Uncertainties in Geometric Parameters}
\label{geompars}

Before computing the uncertainties on the rotation velocities
themselves, we must first determine the uncertainties on each of the
parameters that are used to calculate the rotation velocities: the
center, position angle (PA), inclination, and systemic velocity.  We
used a bootstrap resampling technique to calculate the value and
uncertainty of these parameters.  We constructed 200 bootstrap samples
of the velocity field and ran our tilted-ring modeling routine ({\sc
ringfit}) on each of them to determine the best-fit kinematic center.
We then defined the center of the galaxy to be the median of these 200
measurements, and the systematic uncertainty on the center position to
be the dispersion of the measurements.  We found that the location of
the kinematic center of NGC~2976 is consistent with the optical
nucleus, with an uncertainty of 2\arcsec\ in both $\alpha$ and
$\delta$.  We used the same bootstrap method to measure the kinematic
PA of the galaxy and its uncertainty, finding that the kinematic PA
matches the photometric one and has an uncertainty of 5\deg.  It is
not possible to determine a kinematic inclination angle for NGC~2976
because the rotation curve is too close to solid-body.  However, the
photometric inclination angle is well determined, and we estimate that
the uncertainty is 3\deg.

The uncertainty in the systemic velocity was calculated in a different
way.  The systemic velocity of each ring was left free to vary during
our tilted-ring fitting, resulting in 23 independent measurements.
The overall uncertainty from these measurements is 1.8 km~s$^{-1}$.
Since $v_{sys}$ is always a free parameter, this uncertainty does not
directly factor in to the uncertainty on the rotation curve.

\subsubsection{Uncertainties in Rotation Velocities and Radial Velocities}
\label{dvrot}

Using the measured uncertainties in the center position, PA, and
inclination angle, we calculated the resulting uncertainties on the
rotation velocities and the radial velocities with a Monte Carlo
study.  We generated 1000 random centers, PAs, and inclinations,
assuming a Gaussian distribution for each of the parameters, and ran
{\sc ringfit} with each set of parameters.  The standard deviation of
the 1000 rotation velocities in each ring was defined to be the
systematic error of that rotation velocity, and the systematic errors
in the radial velocities and systemic velocities were calculated in
the same way.  The systematic errors on the rotation curve range from
2.1 km~s$^{-1}$ to 5.5 km~s$^{-1}$.  The sum in quadrature of these
errors and the statistical errors translates to a total uncertainty on
the density profile slope $\alpha$ of 0.09.  We conclude that, even in
the minimum disk case, an NFW or steeper density profile in NGC~2976
is strongly ruled out.

\subsection{Other Systematic Effects}

In some studies, beam-smearing is one of the dominant systematic
effects.  The velocity field of NGC~2976, though, is extremely
well-resolved --- our spatial resolution is less than 100 pc in both
CO and H$\alpha$.  The measured velocity gradient is small ($\sim 50$
km~s$^{-1}$~kpc$^{-1}$) and the constant-density core that we detected
is $\sim50$ resolution elements across, indicating that beam-smearing
is not responsible for this result.  We also calculated the asymmetric
drift correction to the rotation curve.  Its effect is to increase the
maximum disk mass-to-light ratio, and to make the rotation curve
slightly more linear, but the density profile does not change
significantly.

\subsection{Comparing Velocities Derived From Different Tracers}
\label{velcompare}

Some recent studies in the literature have shown that, beam smearing
questions aside, there do not appear to be systematic offsets between
H {\sc i} and H$\alpha$ rotation velocities (e.g., McGaugh, Rubin, \&
de Blok 2001; Marchesini et al. 2002).  With a handful of exceptions,
though, these studies employed longslit H$\alpha$ data, so the
comparisons essentially took place only along the major axis.  In
addition, the spatial and velocity resolution of the H {\sc i} and
H$\alpha$ data were often quite different.

Simon et al. (2003) presented for the first time the data necessary
for a two-dimensional comparison across a dwarf galaxy of the CO and
H$\alpha$ velocity fields.  The angular resolution of the two datasets
is similar (6\arcsec\ and 4\arcsec, respectively), and although the CO
velocity resolution is better by a factor of $\sim6$, the higher
signal-to-noise at H$\alpha$ enables us to measure the velocities with
comparable precision.  We constructed a unique one-to-one mapping
between the two velocity fields.  The mean difference between
$v_{\mbox{{\tiny \mbox{H$\alpha$}}}}$ and $v_{\mbox{{\tiny CO}}}$ is 1
km~s$^{-1}$ and the rms is 5.3 km~s$^{-1}$, with the comparison being
made at 173 independent points.  Similar studies in the Milky Way
found that the dispersion between the velocities of molecular clouds
and the associated H$\alpha$-emitting gas was 4-6 km~s$^{-1}$ (Fich,
Dahl, \& Treffers 1990), so much of the scatter we observe between the
two in NGC~2976 may be intrinsic to the process of H {\sc ii} region
formation rather than caused by observational uncertainties.

This comparison strongly suggests that both the CO and the H$\alpha$
data accurately reflect the gravitational potential of the galaxy, and
neither is significantly compromised by systematic effects.
Extinction, for example, is not distorting the H$\alpha$ velocity
field.  If this result can be shown to hold for a few additional
galaxies, then it will no longer be necessary to observe more than one
velocity field tracer per galaxy.  One further test that should be
carried out is to compare the velocity field from a gaseous
emission-line tracer to that from a stellar absorption-line tracer to
assess the agreement between the stellar and gas kinematics.  Such
studies have been done before (e.g., Mulder 1995; Bottema 1999), but
usually only with long-slit observations.  Obtaining two-dimensional
absorption-line velocity fields is currently observationally
challenging but feasible.

\subsection{Noncircular Motions}
\label{noncirc}

Perhaps the most surprising result from Simon et al. (2003) was the
finding that NGC~2976 contains strong radial motions.  In optical and
near-infrared images, the galaxy is very regular, with no hint of a
bar or any other deviation from axisymmetry that might affect the
kinematics.  The finding that the velocity field is clearly different
from a purely rotating disk was therefore unexpected.  Near the center
of the galaxy, the direction of the maximum velocity gradient is up to
$\sim40\deg$ away from the photometric major axis.

The velocity data can be adequately described by a model in which the
galaxy is undergoing circular rotation, but only if the kinematic PA
declines monotonically from $\sim6\deg$ to $\sim-37\deg$ over the
central kiloparsec.  As mentioned above, though, this position angle
variation is inconsistent with the photometry.  The alternative ---
and more likely --- model is one in which NGC~2976 contains
substantial radial flows.  The origin of such motions, however, is
unclear.

There are several plausible mechanisms for creating radial motions.
An intriguing possibility is that the radial motions could be a result
of the dark matter halo having a triaxial rather than spherical shape.
CDM halos are expected to be moderately triaxial (e.g., Dubinski \&
Carlberg 1991; Warren et al. 1992), and the velocity field of a galaxy
embedded in a triaxial halo would exhibit noncircular motions.
However, since the details of such a velocity field have not yet been
investigated with simulations, we cannot compare our results to
theoretical predictions.  We plan to carry out simulations of the
kinematics of a disk within a triaxial halo to see how the halo shape
influences the gas kinematics and whether the effect is consistent
with the velocity field of NGC~2976.  This could become a new
technique for measuring the triaxiality of galaxy dark matter halos.

\section{The Dark Matter Halo of NGC~4605}
\label{ngc4605}

\begin{figure}[!t]
\plotone{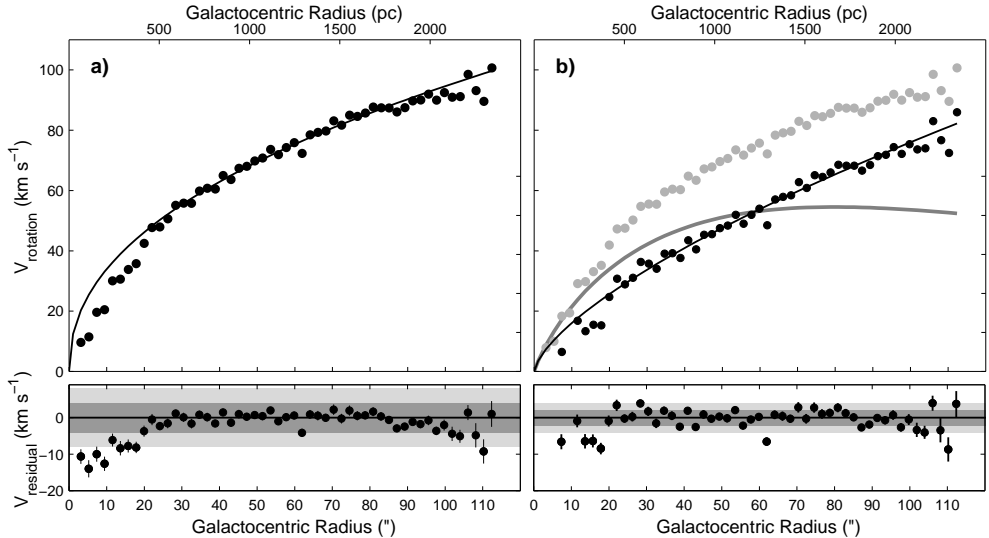}
\caption{(a) Minimum disk rotation curve of NGC~4605.  The black curve
is a $\rho \propto r^{-1.1}$ power law fit, which agrees quite closely
with the rotation curve beyond 25\arcsec.  At smaller radii, though,
it significantly overestimates the rotation velocities.  A separate
inner power law ($\rho \propto r^{-0.4}$) is necessary to fit the
entire rotation curve.  The bottom panel shows the residuals from the
fit, with $1\sigma$ and $2\sigma$ departures represented by the shaded
regions.  Statistical error bars are displayed in the residual plot.
(b) Maximum disk rotation curve of NGC~4605.  The light gray circles
represent the observed rotation curve, and the black circles represent
the rotation curve of the dark matter halo after the stellar disk
contribution (thick gray curve) has been subtracted.  The black curve
is a $\rho \propto r^{-0.65}$ power law that provides a good fit to
the dark matter rotation curve at all radii.
\label{ngc4605rc}}
\end{figure}

NGC~4605 is another Sc dwarf galaxy, located at a distance of 4.3 Mpc.
Like NGC~2976, it appears to be a pure disk system.  We used a
long-slit H$\alpha$ spectrum and a two-dimensional CO velocity field
to measure its rotation curve (Bolatto et al. 2002).  We found that
the galaxy probably contains a maximal disk, and that its dark matter
density profile goes as $\rho \propto r^{-0.65}$ (see
Fig. \ref{ngc4605rc}), intermediate between a core and a cusp.  The
quality of our original observations of NGC~4605 precluded an
investigation of the systematics, but we are in the process of
analyzing new observations to study their impact.

\section{The Dark Matter Halo of NGC~5963}
\label{ngc5963}

NGC~5963 is a more massive ($\sim10^{10} M_{\odot}$) and more distant
($\sim10$ Mpc) galaxy than NGC~2976 and NGC~4605.  Its surface
brightness profiles are more complicated than those galaxies as well,
showing bright emission at the center, surrounded by a transition
region in which the surface brightness falls off extremely rapidly
(scale length $\approx 300$ pc), and then a very low surface
brightness exponential disk extending from 1.5 kpc out to beyond 3
kpc.  The interpretation of this behavior near the center of the
galaxy is not straightforward; the galaxy is definitely not strongly
barred, but we cannot rule out a bulge component.

\subsection{Observations}

We obtained a two-dimensional H$\alpha$ velocity field of NGC~5963
with the WIYN telescope, using the same technique as for NGC~2976.
The spatial resolution of these data is 190 pc and the velocity
resolution is 13 km~s$^{-1}$.  We detected H$\alpha$ emission out to a
radius of $\sim3$ kpc.  We also obtained a two-dimensional CO velocity
field from BIMA.  The CO emission in this galaxy is limited to the
central kiloparsec, but adding the CO data improves our ability to
constrain the innermost part of the density profile.

\subsection{Rotation Curve and Stellar Disk}

\begin{figure}[!t]
\plotone{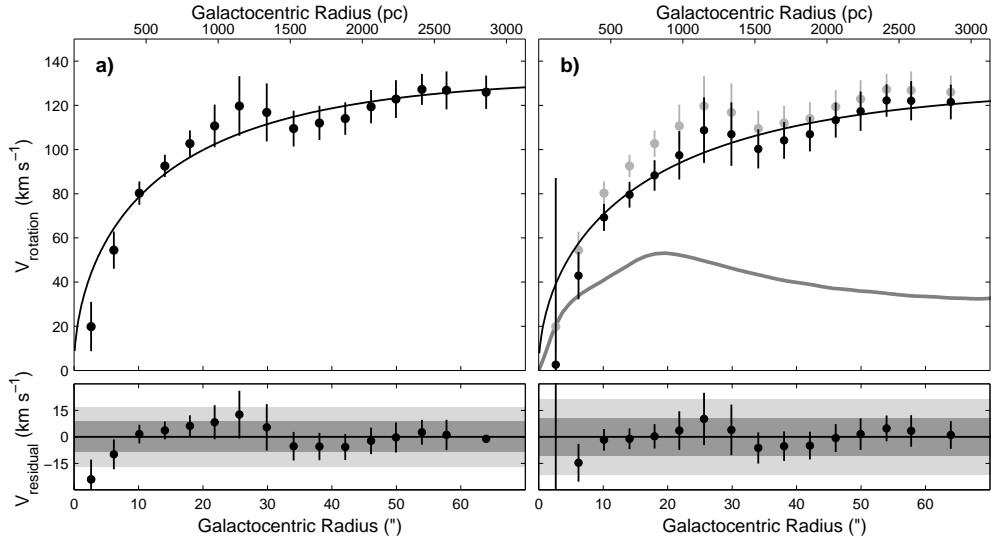}
\caption{(a) Minimum disk rotation curve of NGC~5963.  The black curve
shows an NFW fit to the rotation curve with a concentration of 20.
Residuals from the fit are plotted in the bottom panel.  Unlike the
previous two galaxies, a power law is not a very good fit and the NFW
form fits much more closely.  (b) Maximum disk rotation curve of
NGC~5963.  Even after removing a maximal stellar disk (thick gray
curve; $M_{*}/L_{I} = 0.7 M_{\odot}/L_{\odot I}$), the rotation curve
due to the dark matter halo is well-fit by an NFW potential.
\label{rcfig5963}}
\end{figure}

The combined CO and H$\alpha$ rotation curve of NGC~5963 is displayed
in Figure \ref{rcfig5963}.  It is immediately clear that the density
profile of NGC~5963 has a very different structure from those of
NGC~2976 and NGC~4605.  Instead of a mostly linear increase with
radius, the rotation curve of NGC~5963 has the classic spiral galaxy
shape with a steep inner rise and a nearly flat outer portion.

We perform isophotal fits to the I-band image of NGC~5963 to obtain
the shape of its surface brightness profiles.  We then calculate the
rotation curve due to the stellar disk numerically, as described
before.  The resulting stellar disk rotation curve can be seen in
Figure \ref{rcfig5963}.  The maximum disk limit for this galaxy is set
by the innermost point of the rotation curve.  We derive a maximum
disk mass-to-light ratio of $M_{*}/L_{I} = 0.7 M_{\odot}/L_{\odot I}$.
Even the maximum disk is unable to account for the observed rotation
speed of the galaxy beyond a radius of $\sim6$\arcsec.  The calculated
value of the maximum disk mass-to-light ratio is not unusually low, so
a maximum disk is not required in NGC~5963.  The disk can be less
massive than shown in Figure \ref{rcfig5963}, but since the galaxy is
dark matter-dominated the structure of the dark halo does not depend
strongly on the assumed stellar mass-to-light ratio.

\subsection{Dark Matter Density Profile}

Fitting a single power law to the rotation curve of NGC~5963 does not
produce a very good fit.  The fit is substantially improved by using
separate inner and outer power laws, corresponding to an inner density
profile of $\rho \propto r^{-1.1}$ and a steeper outer density profile
of $\rho \propto r^{-1.4}$.  An even better fit can be obtained by
fitting an NFW rotation curve to the data; the reduced $\chi^{2}$
value of this fit is only 1.1.  Although the concentration parameter
of the halo is not well-constrained, the scale radius is about 3 kpc
and the halo has $v_{200} \approx 90$ km~s$^{-1}$.  An NFW fit can
accurately describe the rotation curve for any stellar mass-to-light
ratio (see Fig.  \ref{rcfig5963}).  This galaxy may be the first
low-mass system found that contains a density profile that matches the
predictions of the numerical simulations.

\section{Conclusions}
\label{conclusions}

We have used high-resolution two-dimensional velocity fields to study
the dark matter density profiles of NGC~2976, NGC~4605, and NGC~5963.
We showed that these three galaxies contain very different density
profiles.  For NGC~2976, we presented one of the most detailed
velocity fields of a dwarf galaxy outside the Local Group.  Even with
these high-quality data, we found that NGC~2976 contains a
constant-density core and an NFW halo is strongly ruled out.  The dark
matter halo of NGC~4605 is intermediate between a constant-density
core and a cusp, and NGC~5963 has a cuspy halo that is almost
perfectly consistent with an NFW profile.

We conclude that some galaxies do not contain central cusps, even when
systematic uncertainties are accounted for as carefully as possible.
We also point out that we find no evidence to support the prediction
that all dark matter halos share a universal shape.  If these results
hold as our sample grows, it will demonstrate that the conflict
between the observations and the simulations is not caused by
systematic problems with the observations.  Instead, more effort may
be needed to investigate what could be missing from the simulations
that would cause them to overestimate density profile slopes.

\acknowledgements{This research was supported by NSF grant
AST-9981308.}

\end{document}